\documentclass{article}
\usepackage{amsmath,graphicx,fullpage,setspace,caption,booktabs,xcolor,hyperref,pdflscape,placeins}
\usepackage[T1]{fontenc}
\usepackage[utf8]{inputenc}
\usepackage{textcomp}
\usepackage{mathptmx}
\ifdefined\pdfminorversion\pdfminorversion=7\fi

\hypersetup{hidelinks,unicode=true}
\title{\textbf{Noise-resilient nanophotonic gyroscope with sub-prad phase resolution}}
\author{\small Yu Tao$^{1,2,3,5}$, Xinhang Li$^{1,5}$, Jingzheng Huang$^{1,2,3,*}$, Zidi Lin$^1$,\\
\small Yuyao Guo$^{1}$, Hongjing Li$^{1,2,3}$, Linjie Zhou$^{1,4}$, Guihua Zeng$^{1,2,3,\ddagger}$\\[5pt]
\scriptsize $^1$State key laboratory of photonics and communications, Shanghai Jiao Tong University, Shanghai 200240, China\\
\scriptsize $^2$Hefei National Laboratory, Hefei 230088, China\\
\scriptsize $^3$Shanghai Research Center for Quantum Sciences, Shanghai 201315, China\\
\scriptsize $^4$SJTU-Pinghu Institute of Intelligent Optoelectronics, Pinghu 314200, China\\
\scriptsize $^5$ These authors contributed equally to this work\\
\scriptsize $^*$jzhuang1983@sjtu.edu.cn; $^\ddagger$ghzeng@sjtu.edu.cn}
\date{}

\begin{document}

\maketitle

\fontsize{10pt}{13pt}\selectfont
{\setstretch{1.0}
\fontsize{10pt}{14pt}\selectfont
\section*{Abstract}
Optical gyroscopes based on the Sagnac effect are the cornerstone of precision orientation and navigation. However, their bulky form factors prevent deployment in emerging mobile and autonomous systems. On nanophotonic platforms, the Sagnac signal plummets under aggressive miniaturization. Consequently, the signal is easily swamped by refractive-index fluctuations, rendering navigation-grade sensitivity within just a few square millimeters a notoriously elusive goal. Here, we demonstrate a noise-resilient nanophotonic optical gyroscope by exploiting a two-chain decoupling architecture to effectively isolate the rotation signal from channel noise. Implemented on a $3~\mathrm{mm}^2$ passive silicon-nitride chip, the proof-of-concept device achieves a bias instability of {$0.79^{\circ}/\mathrm{h}$} and an angle random walk of {$0.0014^{\circ}/\sqrt{\mathrm{h}}$}, representing improvements of 4 and 6 orders of magnitude, respectively, over {a} representative nanophotonic gyroscope {with the same enclosed sensing area~\cite{khial2018nanophotonic}}. In the broader context of integrated optical gyroscopes, our approach bridges the long-standing size--performance gap by two to three orders of magnitude, moving chip-scale devices into a previously inaccessible regime and pointing toward navigation-relevant precision for monolithic microsystems. This architecture further enables {an estimated input-equivalent} phase resolution {of $0.17~\mathrm{prad}$} with general applicability, establishing a foundational framework for the next generation of robust, monolithically integrated photonic sensing systems.
}

\fontsize{10pt}{13pt}\selectfont
\newpage

\section*{Introduction}
Gyroscopes provide direct measurements of angular velocity, the primary dynamical variable that determines how an object changes its orientation~\cite{lawrence2001modern}. As a fundamental category of inertial sensors, they are indispensable across a vast spectrum of applications~\cite{dell2023miniaturization}. In recent years, the rise of emerging microscale platforms, such as insect-scale vehicles~\cite{lu2025insect,shen2024sunlight,farrell2018review}, microrobotic systems~\cite{sui2025untethered,jancik2025nano,jiang2022control}, and wearables~\cite{Chen2025NoiseTolerantHMI,ates2022end}, has intensified the need for gyroscopes that meet stringent size, weight, and power constraints, which conventional architectures can no longer satisfy. Realizing high-performance gyroscopes that are substantially smaller than these platforms, and compatible with existing mass-production processes, is therefore essential for their eventual large-scale deployment, while also opening new regimes of angular sensing that have previously remained inaccessible.

Optical gyroscopes are widely regarded as the most promising route toward realizing high-performance integrated inertial sensors. Their lack of moving mechanical components enables noise suppression at the level of the underlying physical mechanisms, in sharp contrast to micro-electromechanical system (MEMS) gyroscopes, whose vibrating structures introduce multiple intrinsic noise pathways that are difficult to overcome, such as mechanical shocks and random vibrations~\cite{kowaltschek2012lessons}. However, integrated implementations on nanophotonic platforms have long struggled to translate this conceptual advantage into experimental reality. Although the waveguide gyroscope concept was proposed as early as 1983\cite{haavisto1983thin}, progress has been severely constrained by a fundamental physical bottleneck arising from the Sagnac effect itself: the rotation-induced phase shift scales linearly with the enclosed area of the optical path\cite{post1967sagnac}. Consequently, aggressive miniaturization drastically reduces the Sagnac signal to levels that are extremely difficult to detect. While decades of development in fiber-optic gyroscopes have successfully harnessed this effect using long optical fiber coils, establishing them as one of the most commercially mature high-precision gyroscopes~\cite{song2023advanced}, nanophotonic implementations remain limited by this harsh area-scaling law.

Efforts to miniaturize optical gyroscopes have predominantly focused on resonator-based architectures to balance the trade-off between size and performance. These structures confine light within a compact resonator, effectively increasing the interaction length within a small footprint~\cite{bogaerts2012silicon,jiang2020whispering}. Such resonant gyroscopes depart from the traditional interferometric readouts of the Sagnac phase~{\cite{dell2023miniaturization, lefevre1993fiber}} and instead extract angular velocity from rotation-induced shifts in the resonance frequency. The pursuit of narrower resonances and higher frequency discrimination has driven the field to explore the limits of both passive ultra-high quality factor(Q) microcavities~\cite{liang2017resonant,liu2025chip} and active gain-based schemes such as Brillouin gyroscopes~{\cite{li2017microresonator,lai2020earth,ge2026integrated}}, aiming to push frequency resolution beyond what planar waveguide resonators can offer~\cite{suzuki2000monolithically,wang2015resonator}. {Beyond improvements in frequency resolution, noise-resilient readout schemes such as reciprocal sensitivity enhancement~\cite{khial2018nanophotonic} and weak-value amplification~\cite{huang2021rotation,song2021enhanced,yanik2025development} offer another route to improved gyroscope performance.} However{, for approaches that infer rotation from resonance-frequency changes}, a critical issue has received far less attention: their response is intrinsically coupled to refractive-index fluctuations. This coupling becomes particularly pronounced at the nanophotonic scale and fundamentally limits the achievable stability, preventing these devices from realizing the full performance potential of resonant enhancement.

Here, we introduce a new two-chain decoupling gyroscope architecture to overcome the fundamental refractive-index--coupling bottleneck of resonant readouts and the challenge of angular velocity extraction under extremely low signal-to-noise ratio conditions. We realize this architecture on a monolithic silicon nitride platform{, using passive optical transduction and static thermo-optic tuning for open-loop angular-rate readout} without {dynamic} modulation. With an enclosed Sagnac area of $0.8~\mathrm{mm}^2$ in a $3~\mathrm{mm}^2$ footprint, the device achieves a bias instability of {$0.79^{\circ}\mathrm{/h}$} and an angle random walk of {$0.0014^{\circ}/\sqrt{\mathrm{h}}$ with an external narrow-linewidth laser}, reaching a previously inaccessible regime of size--performance capability for integrated optical gyroscopes. Crucially, {dynamic measurements yield} an {estimated input-equivalent} phase resolution of {$0.17~\mathrm{prad}$, extending the significance of this performance to optical phase sensing}. {We further demonstrate operation with an on-chip self-injection-locked (SIL) laser}, {achieving a bias instability of \(5.85^{\circ}/\mathrm{h}\) and an angle random walk of \(0.00439^{\circ}/\sqrt{\mathrm{h}}\), supporting} the {development of compact integrated gyroscope systems}. By unlocking this regime, our work can extend beyond inertial sensing and place integrated nanophotonic metrology on a new performance frontier.
\section*{Results}\label{results}

\subsection*{Two-chain decoupling architecture}
Fig. 1a illustrates the architecture of our noise-resilient nanophotonic gyroscope, defined by a two-chain decoupling framework that structurally isolates signal generation from channel noise. The signal-generation chain encodes rotation into a resonantly enhanced, refractive-index-independent nonreciprocal differential phase within a symmetric dual-resonator interferometric structure. Subsequently, during the readout process, this pristine signal becomes inevitably mixed with systemic disturbances outside the generation core, which we collectively term the channel noise. This noise typically manifests as intensity fluctuations that are indistinguishable from the rotation signal. To resolve this ambiguity, we strategically introduce a static symmetry-breaking phase bias. This controlled asymmetry shifts the operating point, enabling the readout to selectively extract the nonreciprocal phase while effectively suppressing the channel noise.

\begin{figure*}[tbp]
    \centering
    \includegraphics[width=0.9\linewidth]{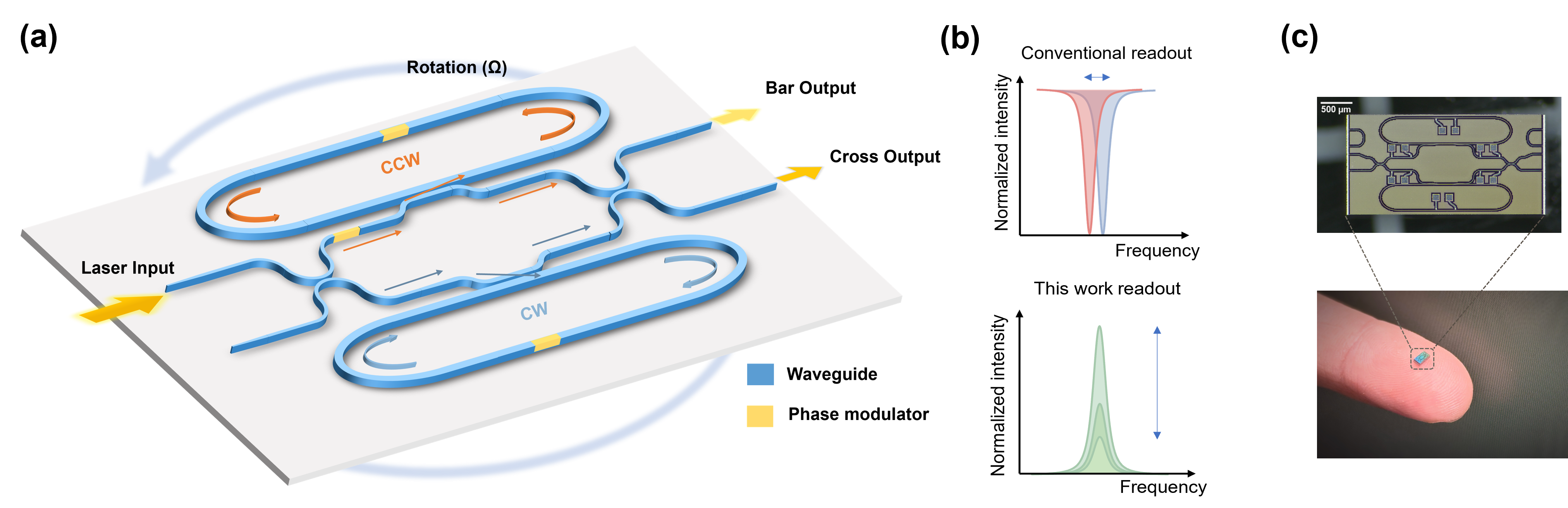}
    \caption{\textbf{Architecture and readout principle of the noise-decoupled nanophotonic gyroscope.}
(a)  Schematic of the proposed architecture. The silicon nitride (Si$_3$N$_4$) waveguides are in blue. The phase modulators are in yellow; the arm modulator provides the symmetry-breaking bias, while the resonator pair is tuned to align their initial resonance condition.
({b}) Conceptual comparison between conventional frequency-tracking readout and the proposed intensity-based readout. In the presented architecture, rotation information is mapped onto intensity variations, fundamentally relaxing the requirements on frequency tracking and source coherence.
(c) Photographs of the fabricated chip after dicing, including a macroscopic image with a fingertip for intuitive size reference and a zoomed-in view showing the on-chip photonic structures. }
    \label{fig:1}
\end{figure*}

Conventional integrated resonant gyroscopes typically infer the rotation rate $\Omega$ by tracking the resonance-frequency splitting $\Delta f_{\mathrm{Sagnac}}$ between clockwise (CW) and counter-clockwise (CCW) modes (Fig. 1b, top). Although the Sagnac phase $\Delta\phi_{\mathrm{Sagnac}}$ itself is refractive-index independent \cite{lefevre1993fiber}, once rotation perturbs the CW/CCW modes from degeneracy to detuning, reestablishing the resonance condition reconstructs the resonance frequencies and reintroduces the refractive index $n$ into the frequency-splitting relation \cite{malykin2014sagnac}. As a result, in demodulation relations based on $\Delta f_{\mathrm{Sagnac}}$ (e.g., $\Delta f_{\mathrm{Sagnac}}\propto \Omega/n$), a refractive-index fluctuation $\delta n$ produces a proportional fluctuation in $\Delta f_{\mathrm{Sagnac}}$, yielding a multiplicative error term that scales with $\Omega$. Unlike additive noise, which sets a fixed noise floor, this multiplicative coupling grows with the measurand, and is difficult to suppress by simple averaging or filtering, and can manifest as an effective noise-induced drift that degrades bias stability and long-term performance \cite{volpe2016effective}. In addition, frequency-tracking schemes are intrinsically susceptible to frequency-domain disturbances such as laser frequency noise.

To eliminate {explicit} multiplicative coupling {to the refractive index} at the signal-generation stage, we propose a {signal-encoding} scheme that {directly} extracts the {nonreciprocal} Sagnac phase. We employ a pair of micro-resonators to spatially separate the CW and CCW optical paths, placing them as phase delay elements in the two arms of an equal-arm Mach-Zehnder interferometer (MZI). In the static state, both rings are biased to resonance, balancing the interferometer. Under rotation, the resonators act as sensitive phase transducers rather than simple spectral filters. The rotation-induced Sagnac effect introduces a nonreciprocal phase shift between the upper and lower arms. Leveraging the steep phase response near resonance, even a minute Sagnac perturbation is amplified into a significant macroscopic phase difference. The MZI then converts this differential phase into an intensity variation at the output port. Consequently, the system response shifts from a resonance-frequency readout to an intensity-based readout (Fig. 1b, bottom{). Under near-resonant operation with approximately stable coupling and loss parameters, the indirect influence of refractive-index variations on the phase-enhancement factor can be neglected (Methods}). This mechanism is experimentally validated by the spectra in Extended Data Fig.~\hyperref[fig:ED1]{1}, which show that phase perturbations that would normally manifest as a horizontal resonance-frequency shift in standard resonator readout are instead converted into a vertical change in the resonance feature amplitude in the intensity--frequency map. Details on the linear dynamic range are discussed in the Supplementary Information.

Ideally, the proposed symmetric architecture allows for the extraction of the clean rotation response while providing perfect common-mode rejection. However, in realistic photonic chips, inevitable fabrication imperfections disrupt this symmetry, causing complex and broadband channel noise to leak back into the signal readout. To address this, we further protect the signal from this noise by leveraging post-selection and weak value amplification, which is a metrological protocol rooted in quantum theory~\cite{aharonov1988result}. As detailed in the Methods, by employing a minute, tunable phase bias, we define an intrinsically differential observable in the measurement operator space rather than the signal space. This biasing operation configures the architecture as a post-selection filter: it projects the system state (the superposition of optical paths) onto a post-selection state nearly orthogonal to the input, which is physically accessible at the bar port. This projection selectively enhances the coupling to physics that break time-reversal symmetry, while strongly suppressing the impact of common-mode disturbances. Notably, this operator-space manipulation realizes the essence of our two-chain decoupling framework: it renders the signal chain effectively distinguishable from the channel noise based on their distinct symmetry properties. While non-reciprocal noise sources remain isomorphic to the signal and thus indistinguishable, the noise budget in dielectric platforms is overwhelmingly dominated by the reciprocal disturbances we effectively suppress. This mechanism {suppresses} noise contributions {originating} outside the signal-generation {chain}, consistent with the {reduced drift and noise background} observed in subsequent experiments.

\fontsize{10pt}{14.4pt}\selectfont
\subsection*{On-chip implementation and performance characterization}

\begingroup\color{black}
The proposed gyroscope architecture was implemented on a silicon nitride (Si$_3$N$_4$) nanophotonic platform. The chip occupies approximately $3~\mathrm{mm}^2$, and each ring resonator encloses an area of $0.8~\mathrm{mm}^2$. We mounted the testing system horizontally on a single-axis motorized rotation stage to establish its response to known rotation inputs and characterize its noise under zero-input conditions.

\endgroup

\begin{figure}[tbp]
\centering
\captionsetup{font=small,labelfont=bf}
\includegraphics[width=1\linewidth]{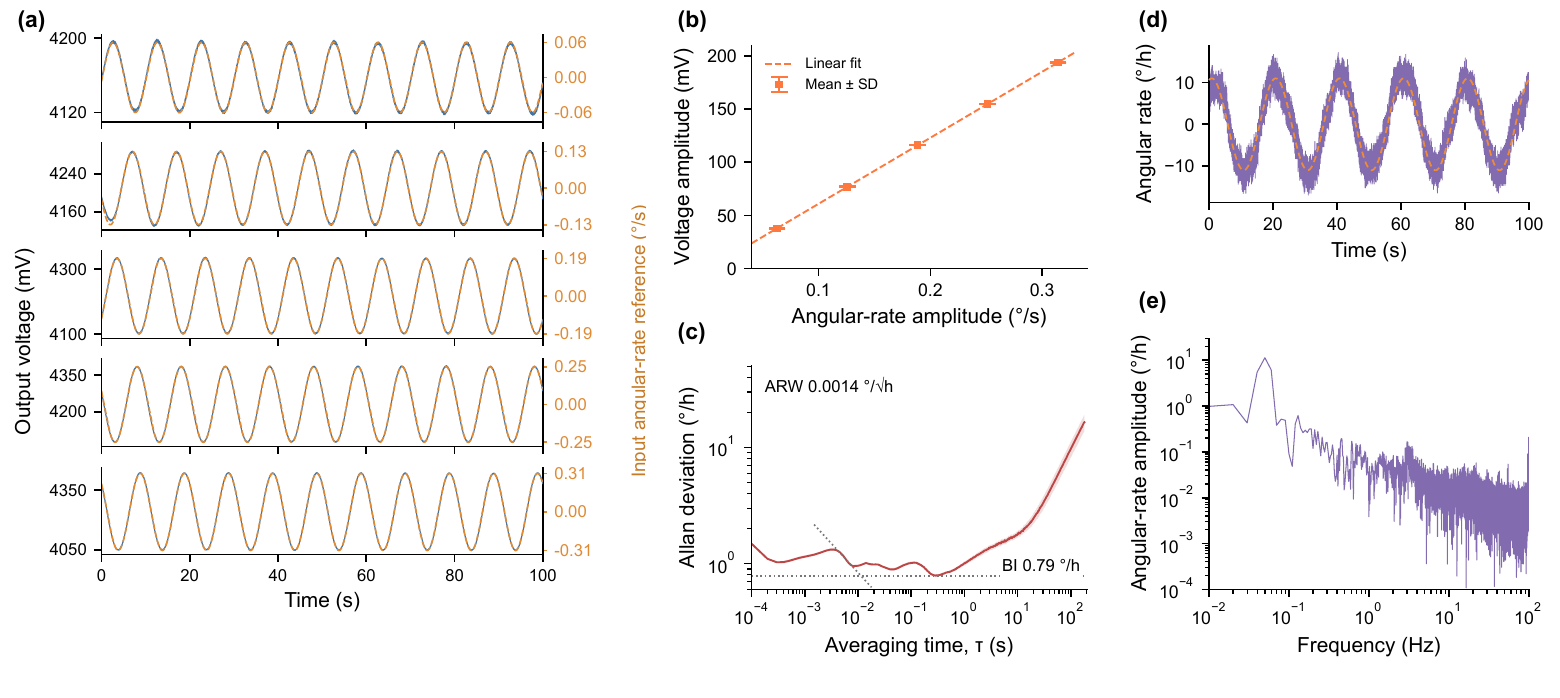}
\caption{{\textbf{Dynamic and static performance and the lowest applied angular-rate input.}
(a) Measured Bar-port voltage responses (left axes) and set angular-rate input references (right axes) at $0.1~\mathrm{Hz}$ for five angular-displacement amplitudes from $0.1^\circ$ to $0.5^\circ$ (top to bottom). Each trace shows the first $100~\mathrm{s}$ of its $500~\mathrm{s}$ record, without additional averaging or smoothing. Orange dashed references use the set input amplitudes, with phases aligned to the output fits.
(b) Physical response calibration. Error bars represent the sample standard deviations of the amplitudes from five consecutive \(100~\mathrm{s}\) intervals; the dashed line is the linear fit.
(c) Allan deviation calculated from the calibrated two-hour static record. Shading indicates pointwise $95\%$ confidence intervals; the sloped dotted line is the ARW fit reference, and the horizontal dotted line marks the minimum of $0.79^\circ/\mathrm{h}$.
(d) A $100~\mathrm{s}$ response to a sinusoidal angular-displacement input at $0.05~\mathrm{Hz}$ with a set peak amplitude of $0.01^\circ$, corresponding to an angular-rate peak amplitude of $11.31^\circ/\mathrm{h}$. The dashed line is the sinusoidal fit.
(e) One-sided amplitude spectrum of the same interval, calculated using a Hann window. The bin spacing is \(0.01~\mathrm{Hz}\), and the equivalent noise bandwidth is \(0.015~\mathrm{Hz}\).}}
\label{fig:main}
\end{figure}

\begingroup\color{black}
An external single-frequency laser was coupled into one input of the $2\times2$ Mach--Zehnder interferometer (MZI). A small static phase bias placed the Bar port near the dark-port condition, enabling weak-value-amplification (WVA) readout with near-orthogonal post-selection. The following dynamic-response and static-noise measurements were performed in this fixed readout configuration.

At this operating point, rotation inputs were directly resolved in the raw voltage output without digital filtering. At $0.1~\mathrm{Hz}$, five angular-displacement amplitudes from $0.1^\circ$ to $0.5^\circ$ produced clear periodic responses, corresponding to angular-rate peak amplitudes of approximately $0.06$--$0.31^\circ/\mathrm{s}$ (Fig.~\ref{fig:main}a). The first $100~\mathrm{s}$ of each record are shown in the main figure; the complete $500~\mathrm{s}$ records are provided in Extended Data Fig.~\ref{fig:supp-dynamic-500s}.

The five responses established a common voltage-to-angular-rate calibration. Calibration points and error bars are the means and sample standard deviations of response amplitudes from five consecutive $100~\mathrm{s}$ intervals in each record. The mean amplitudes varied linearly with input angular rate, yielding a scale factor of $621.17~\mathrm{mV}/(^\circ/\mathrm{s})$ and $R^2=0.999995$ (Fig.~\ref{fig:main}b). At each fixed input amplitude, the interval amplitudes showed little variation, with coefficients of variation (CVs) ranging from \(0.020\%\) to \(0.412\%\). The angular-rate conversion coefficients obtained from calibrations at different input amplitudes were also consistent, with a CV of \(0.110\%\). These results support the short-term repeatability of the response amplitude at fixed input and the consistency of the cross-amplitude calibration across time intervals (Extended Data Fig.~\ref{fig:ed-calibration}; Supplementary Information). This calibration was used to convert static measurements acquired in the matching configuration into angular rate.

The readout also resolved the smallest rotation input that the stage could be set under the present conditions. For a sinusoidal angular-displacement input at $f=0.05~\mathrm{Hz}$ with a set peak amplitude of $\theta_0=0.01^\circ$, the corresponding angular-rate peak amplitude is $\Omega_0=2\pi f\theta_0=11.31^\circ/\mathrm{h}$. The periodic response remained clearly resolved in both the time and frequency domains (Fig.~\ref{fig:main}d,e). This input amplitude was limited by the stage's setting capability. For a single-turn area of $0.8~\mathrm{mm}^2$ and a wavelength of $1550~\mathrm{nm}$, it corresponds to a Sagnac phase amplitude of $2.37~\mathrm{prad}$. A conservative noise estimate based on the mean power spectral density in a band adjacent to the signal gave a spectral power signal-to-noise ratio (SNR) of $22.91~\mathrm{dB}$. Assuming locally linear response and an unchanged noise floor, the estimated input-equivalent phase resolution at SNR$=1$ is $0.17~\mathrm{prad}$ for a $100~\mathrm{s}$ observation time and an equivalent noise bandwidth of $0.015~\mathrm{Hz}$.

\endgroup

\begingroup\color{black}
Whereas known rotation inputs establish the quantitative response, zero-input measurements characterize the static noise of the readout. After conversion with the matching scale factor, the Allan deviation of a two-hour record reached a minimum of $0.79^\circ/\mathrm{h}$ at $\tau=0.29~\mathrm{s}$, reported here as BI (Fig.~\ref{fig:main}c). A free-exponent fit over $4$--$8~\mathrm{ms}$ yielded a slope of approximately $-0.5$; fixing the exponent at $-1/2$ gave an ARW of $0.0014^\circ/\sqrt{\mathrm{h}}$.

Using the same Sagnac conversion, this BI corresponds to an input-equivalent phase fluctuation of approximately $0.17~\mathrm{prad}$, close to the input-equivalent phase resolution of $0.17~\mathrm{prad}$ estimated from the SNR of the dynamic measurement above. The two values are obtained from the minimum static Allan deviation and an extrapolation of the dynamic signal to unity SNR, respectively. Although they correspond to different timescales and criteria, both yield sub-prad input-equivalent phase metrics, supporting the architecture's sub-prad input-equivalent phase readout capability through static stability and dynamic signal distinguishability.

\endgroup

\begingroup\color{black}
\subsection*{Noise suppression in dual-port readout}

\endgroup

\begin{figure}[htbp]
\centering
\captionsetup{font=small,labelfont=bf}
\includegraphics[width=0.85\linewidth]{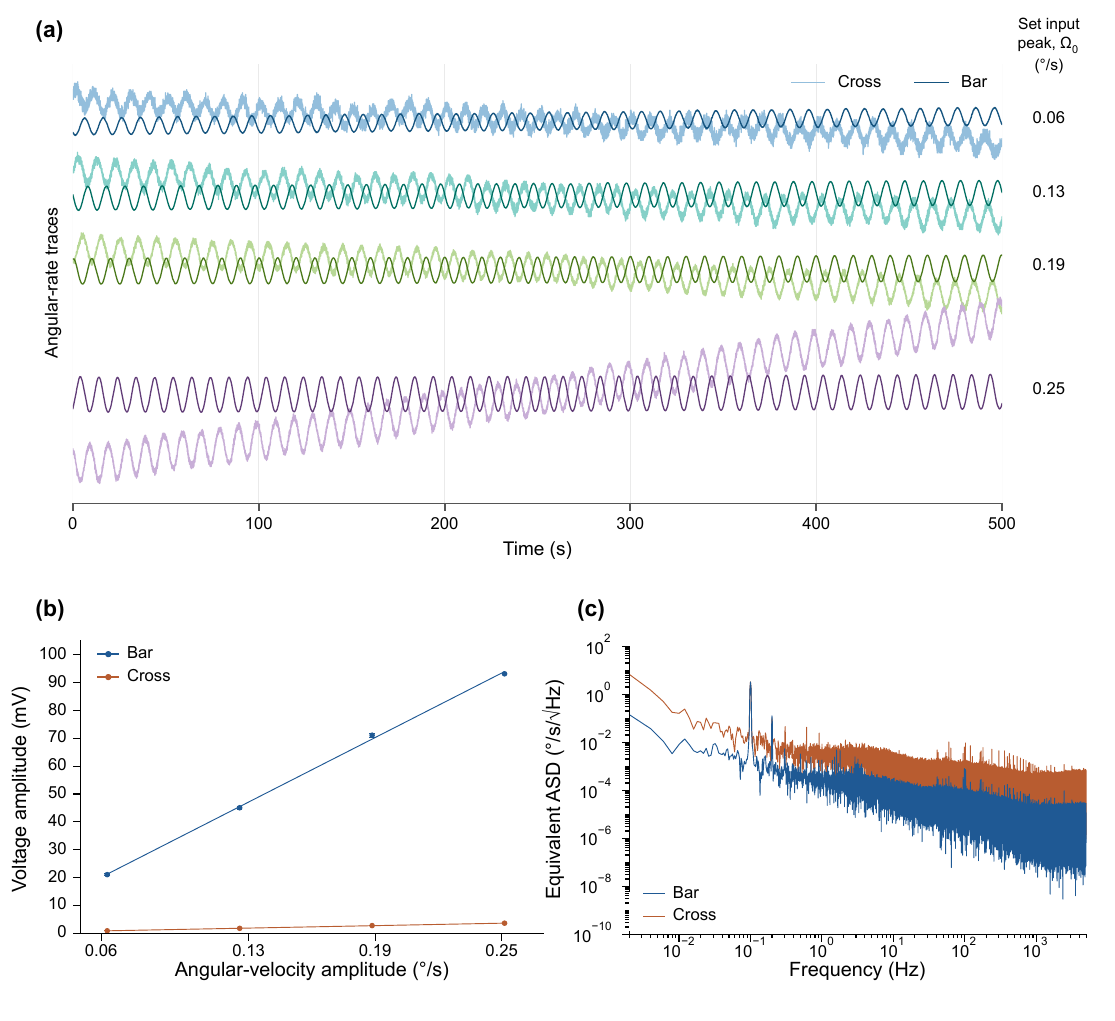}
\caption{{\textbf{Simultaneous comparison of the Bar (WVA) and Cross (non-WVA) ports.}
(a) Simultaneous responses at approximately $0.06$, $0.13$, $0.19$ and $0.25^\circ/\mathrm{s}$, vertically offset for display. For display, the signals from both ports are multiplied by $2$ in the first record and by $1.4$ in the second record. This scaling is not applied in subsequent calibration, drift statistics or spectral calculations.
Values on the right indicate the set input angular-rate peak amplitudes.
(b) Separate physical response calibrations for the two ports. Points are the mean amplitudes from five consecutive $100~\mathrm{s}$ intervals; error bars are sample standard deviations, and solid lines are linear fits.
(c) Input-referred ASD for a peak input angular rate of approximately $0.25^\circ/\mathrm{s}$, calculated from the full $500~\mathrm{s}$ record using a periodic Hann window. The sampling rate is $10~\mathrm{kHz}$, the bin spacing is $0.002~\mathrm{Hz}$, and the equivalent noise bandwidth is $0.003~\mathrm{Hz}$. No segment averaging, smoothing, linear detrending or subtraction of the rotation signal is applied.}}
\label{fig:ports}
\end{figure}

\begingroup\color{black}
The dynamic and static measurements above established the architecture's ability to read out weak rotation. To further examine the influence of background disturbances, we measured the two complementary output ports of the same architecture simultaneously under a common rotation input and experimental environment. At the selected bias, the Bar port provides WVA readout and the Cross port serves as the non-WVA reference. Both ports share the symmetric dual-resonator structure, allowing the different responses of the biased readout to rotation signals and background disturbances to be examined.

The two outputs were acquired simultaneously at $0.1~\mathrm{Hz}$ and four angular-rate amplitudes. Each experiment was accompanied by a calibration under its corresponding operating conditions. In this measurement, separate scale factors for Bar and Cross converted the output voltages into input-referred angular rates, allowing drift and noise to be compared in the same physical units (Fig.~\ref{fig:ports}b). Detectors of the same model and gain setting were used. Fixed attenuation at Cross prevented detector saturation.

The simultaneous measurements first revealed a difference in slow drift. Both ports retained the periodic rotation response, but the Bar-port baseline was more stable (Fig.~\ref{fig:ports}a). To quantify slow drift, we extracted one baseline value per complete rotation cycle and converted these values to input-referred angular rate. The standard deviations of these cycle-wise baseline values at the Cross port were approximately $4$, $8$, $16$ and $48$ times those at the Bar port in the four records, respectively. These results indicate smaller slow drift in the Bar-port angular-rate readout.

This improvement was also evident in the spectral noise floor. At $0.25^\circ/\mathrm{s}$, both ports retained the rotation-signal peak at $0.1~\mathrm{Hz}$, whereas Bar exhibited a lower input-referred amplitude spectral density (ASD) background over a broad frequency range (Fig.~\ref{fig:ports}c). The concurrent reductions in slow drift and broadband noise background show that the biased readout of this architecture improves the distinguishability of rotation signals from background disturbances under the technical-noise conditions of this experiment.

\endgroup

\begingroup\color{black}
\subsection*{Operation with a chip-scale SIL source}

\endgroup

\begin{figure}[tbp]
\centering
\captionsetup{font=small,labelfont=bf}
\includegraphics[width=1\linewidth]{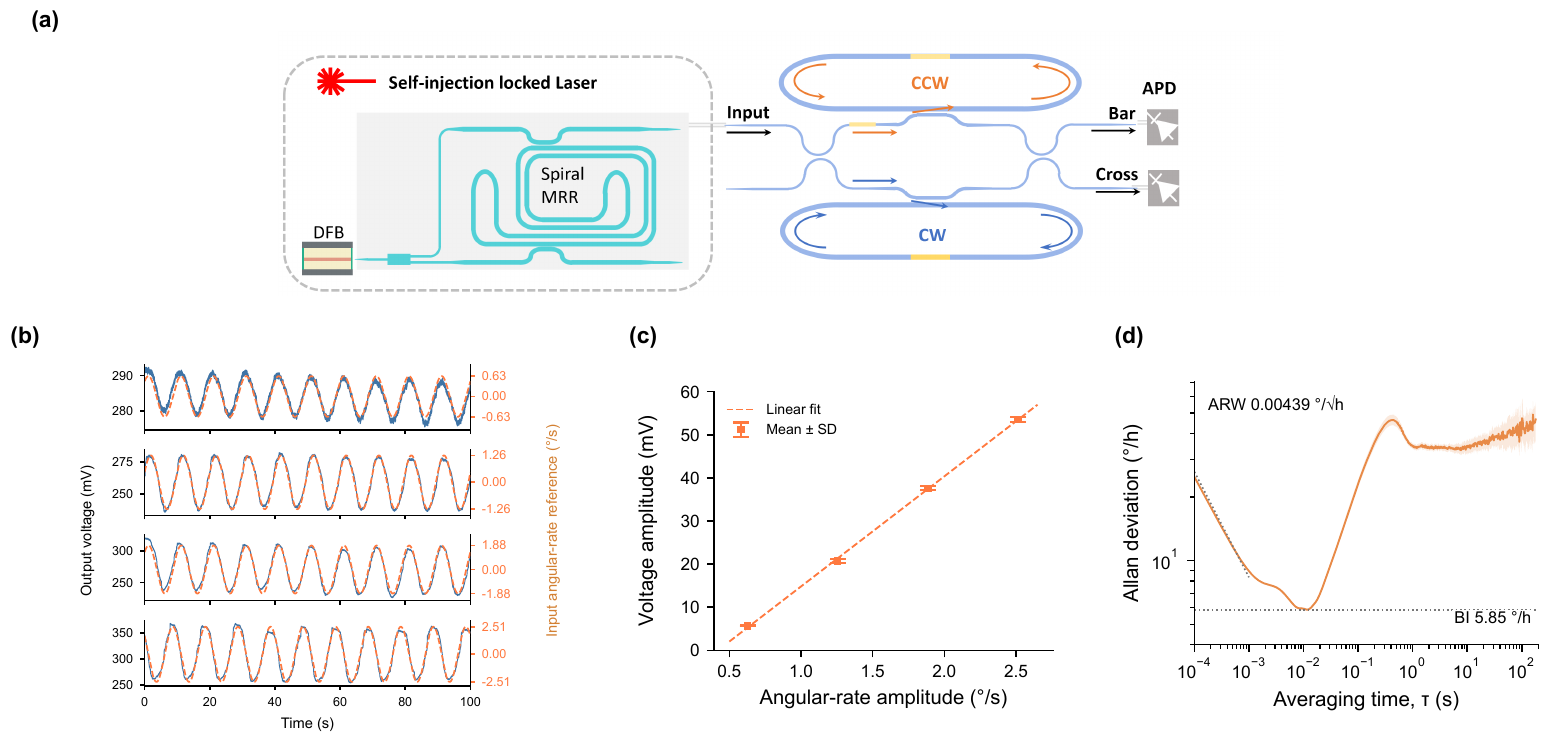}
\caption{{\textbf{Dynamic and static gyroscope operation with a chip-scale SIL source.}
(a) Experimental configuration with the chip-scale SIL source supplying the gyroscope sensing chip.
(b) Voltage responses at $0.1~\mathrm{Hz}$ for angular-displacement amplitudes of $1^\circ$, $2^\circ$, $3^\circ$ and $4^\circ$. Left axes show measured voltage; orange dashed traces and right axes provide the set angular-rate input references. Each displayed interval spans $100~\mathrm{s}$.
(c) Physical voltage-response calibration for the four input conditions. Each $100~\mathrm{s}$ record is divided into five consecutive $20~\mathrm{s}$ intervals. Points and error bars are the means and sample standard deviations of the response amplitudes obtained by fitting each interval at the known driving frequency using a sinusoid with a linear baseline; the dashed line is a linear fit.
(d) Allan deviation from the calibrated two-hour static record. Shading indicates pointwise $95\%$ confidence intervals. The horizontal dotted line marks the minimum Allan deviation, and the sloped dotted line is the estimated ARW reference, using the same plotting conventions as Fig.~\ref{fig:main}c.}}
\label{fig:SIL}
\end{figure}

\begingroup\color{black}
Moving from an external narrow-linewidth laser to a chip-scale source requires both miniaturization and stable optical output. To assess the effect of this transition on gyroscope operation, we replaced the external single-frequency laser with a chip-scale narrow-linewidth SIL laser and characterized the gyroscope's dynamic response and static noise. The source comprises a commercial distributed-feedback (DFB) laser and a low-loss Si$_3$N$_4$ feedback circuit incorporating a high-$Q$ ($Q \sim 10^{6}$) microring resonator (MRR) embedded in a Sagnac loop. In the SIL state, the integrated laser achieved a side-mode suppression ratio (SMSR) exceeding 55~dB and an intrinsic linewidth of approximately 20~Hz. The laser output was connected to the gyroscope sensing chip, as shown in Fig.~\ref{fig:SIL}a. The source and sensing chip were temperature controlled independently, while the gyroscope retained fixed-bias, open-loop Bar-port WVA readout.

With the SIL laser, the device exhibited clear periodic responses to four sinusoidal inputs at $0.1~\mathrm{Hz}$ and angular-displacement amplitudes from $1^\circ$ to $4^\circ$ (Fig.~\ref{fig:SIL}b). Calibration under this configuration yielded a scale factor of $25.54~\mathrm{mV}/(^\circ/\mathrm{s})$, with the response amplitude increasing linearly with input angular rate and $R^2=0.99957$ (Fig.~\ref{fig:SIL}c). Applying the corresponding calibration to a two-hour zero-input record gave an ARW of $0.00439^\circ/\sqrt{\mathrm{h}}$ and a BI of $5.85^\circ/\mathrm{h}$ (Fig.~\ref{fig:SIL}d).

These results demonstrate joint operation of the chip-scale SIL source and gyroscope sensing chip, providing an experimental basis for further integration of the source and sensing unit towards a fully integrated photonic gyroscope.

\endgroup

\subsection*{Size--performance comparison}

\begin{figure}[tbp]
\centering
\captionsetup{font=small,labelfont=bf}
\includegraphics[width=1\linewidth]{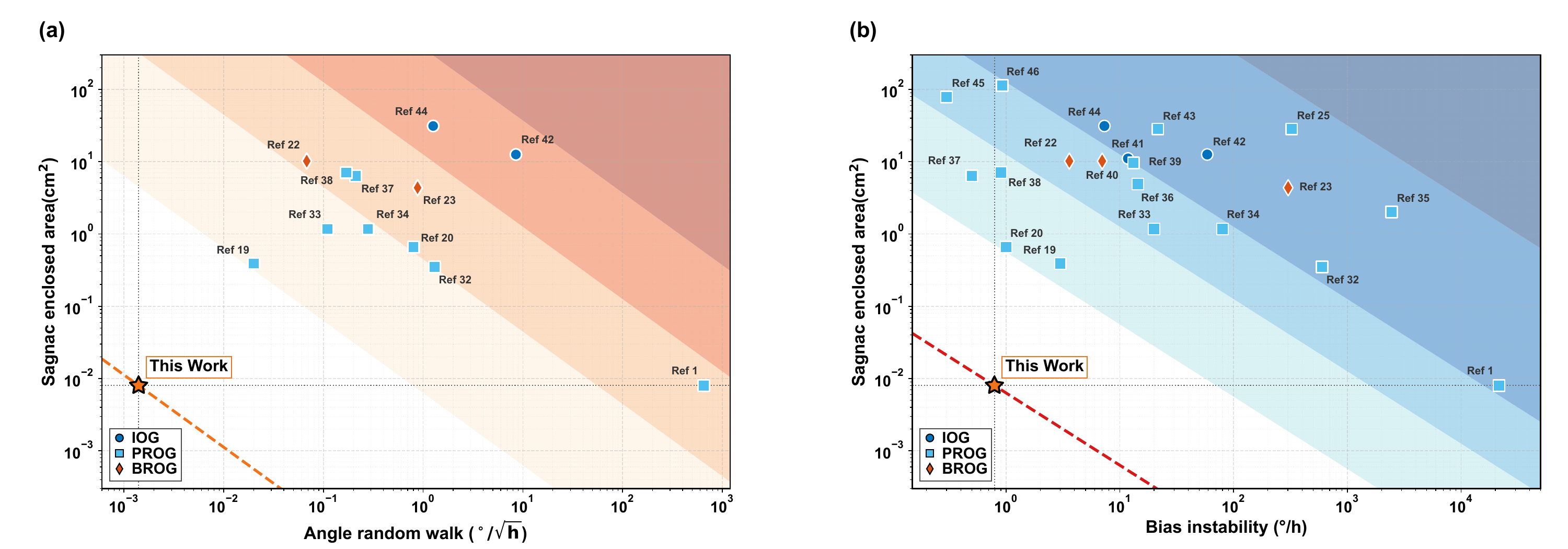}
\caption{{\textbf{Size--performance benchmarking of integrated optical gyroscopes.} (a) Comparison of angular random walk (ARW). (b) Comparison of bias instability (BI). The vertical axis is the geometrical area enclosed by the sensing resonator or coil, characterizing the scale of the sensing element.}}
\label{fig:4}
\end{figure}

\begingroup\color{black}
Figure~\ref{fig:4} benchmarks representative integrated optical gyroscopes in a unified size--performance space, with data sources and metric-extraction procedures detailed in Extended Data Table~1. We use the geometrical area enclosed by the sensing resonator or waveguide coil to characterize the scale of the sensing element, and list ARW and BI values according to the literature reports and documented conversions or extraction procedures. For our device, this area is $0.8~\mathrm{mm}^2$ enclosed by an individual ring, distinct from the total chip footprint of approximately $3~\mathrm{mm}^2$. Literature data broadly follow a direction with a slope of approximately $-1$ in the log--log representation, reflecting an overall size--performance trade-off with platform- and implementation-dependent scatter. The diagonal shaded bands are positioned with reference to this distribution and represent intervals of constant area--performance product, providing a visual comparison of the size--performance combinations achieved by different devices.

Within this comparison, our work achieves an ARW of $0.0014^\circ/\sqrt{\mathrm{h}}$ and a BI of $0.79^\circ/\mathrm{h}$ at a sensing area of $0.8~\mathrm{mm}^2$. The ARW meets the navigation-grade ARW criterion ($<0.005^\circ/\sqrt{\mathrm{h}}$)\cite{dell2023miniaturization}, placing the device at the leading edge of integrated implementations at comparable scales. Relative to a representative nanophotonic gyroscope with the same sensing area\cite{khial2018nanophotonic}, ARW and BI improve by approximately six and four orders of magnitude, respectively. The ARW extracted from the short-time white-noise regime captures the architecture's low-noise readout capability, while BI further demonstrates static stability at the nanophotonic scale. These performance metrics are obtained at fixed bias in a fully open-loop configuration, without active modulation, digital filtering or algorithmic denoising, highlighting the role of structural design rather than complex control in the performance improvement. Collectively, Fig.~\ref{fig:4} demonstrates a breakthrough in the size--performance boundary of nanophotonic gyroscopes: simultaneous, multi-order-of-magnitude improvements in noise floor and stability at an exceptionally small sensing area, opening a new route towards high-precision, fully integrated photonic inertial sensing.

\endgroup

\section*{Discussion and outlook}
By employing the two-chain decoupling architecture that structurally isolates rotational signals from noise channels at the physical level, we realize a bias instability of {$0.79^{\circ}/\text{h}$} and an angle random walk of {$0.0014^{\circ}/\sqrt{\text{h}}$} on a compact $\sim 3~\mathrm{mm}^2$ SiN platform. The performance leap is driven by architectural redesign rather than stringent device parameters, marking a significant milestone in the development of nanophotonic gyroscopes. Practically, our architecture is readily implementable and foundry-compatible. Its architecture-driven noise resilience and generalizability lower the barrier for widespread adoption and pave the way for scalable mass production. Additionally, the straightforward readout and angular rate extraction scheme further minimize the requirements for external digital circuits and computational power. Collectively, these advantages translate into meaningful SWaP-C (size, weight, power, and cost) reductions, which not only extend the operational endurance of micro-platforms but also unlock the potential for ubiquitous inertial sensing in resource-constrained environments.

To further improve long-term stability, future efforts can systematically advance thermal management, environmental isolation (e.g., through improved packaging and temperature-control strategies), and error modeling, forming a complementary framework alongside the proposed architecture. These measures {also complement} the {chip-scale SIL-source operation demonstrated here and provide a route} to {improving} long-term {stability as} the source and {sensing unit are further integrated}. In addressing these practical challenges, the extensive heritage of MEMS gyroscopes in packaging and compensation provides a valuable set of established technologies and methodologies.

Beyond inertial sensing, the sub-prad {input-equivalent} phase {performance demonstrated} by {static measurements and estimated from dynamic measurements with} this two-chain decoupling framework has profound implications for the broader landscape of optical metrology. This suggests that the performance potential of microresonators remains largely untapped, implying that systematic exploration could unlock sensitivity levels once deemed unachievable in integrated platforms. In this sense, our work not only provides renewed confidence and direction for nanophotonic gyroscope research but also establishes a foundational architecture for the next generation of robust and monolithically integrated photonic sensing systems.

\begingroup\bibliography{bibliography}\endgroup

\begin{thebibliography}{10}
\expandafter\ifx\csname url\endcsname\relax
  \def\url#1{\texttt{#1}}\fi
\expandafter\ifx\csname urlprefix\endcsname\relax\def\urlprefix{URL }\fi
\providecommand{\bibinfo}[2]{#2}
\providecommand{\eprint}[2][]{\url{#2}}

\bibitem{khial2018nanophotonic}
\bibinfo{author}{Khial, P.~P.}, \bibinfo{author}{White, A.~D.} \&
  \bibinfo{author}{Hajimiri, A.}
\newblock \bibinfo{title}{Nanophotonic optical gyroscope with reciprocal
  sensitivity enhancement}.
\newblock \emph{\bibinfo{journal}{Nature Photonics}}
  \textbf{\bibinfo{volume}{12}}, \bibinfo{pages}{671--675}
  (\bibinfo{year}{2018}).

\bibitem{lawrence2001modern}
\bibinfo{author}{Lawrence, A.}
\newblock \emph{\bibinfo{title}{Modern inertial technology: navigation,
  guidance, and control}} (\bibinfo{publisher}{Springer Science \& Business
  Media}, \bibinfo{year}{2001}).

\bibitem{dell2023miniaturization}
\bibinfo{author}{Dell'Olio, F.}, \bibinfo{author}{Natale, T.},
  \bibinfo{author}{Wang, Y.-C.} \& \bibinfo{author}{Hung, Y.-J.}
\newblock \bibinfo{title}{Miniaturization of interferometric optical
  gyroscopes: a review}.
\newblock \emph{\bibinfo{journal}{IEEE Sensors Journal}}
  \textbf{\bibinfo{volume}{23}}, \bibinfo{pages}{29948--29968}
  (\bibinfo{year}{2023}).

\bibitem{lu2025insect}
\bibinfo{author}{Lu, X.} \emph{et~al.}
\newblock \bibinfo{title}{An insect-scale flapping-wing micro aerial vehicle
  inspired by tumblers capable of uncontrolled self-stabilizing flying}.
\newblock \emph{\bibinfo{journal}{Research}} \textbf{\bibinfo{volume}{8}},
  \bibinfo{pages}{0787} (\bibinfo{year}{2025}).

\bibitem{shen2024sunlight}
\bibinfo{author}{Shen, W.} \emph{et~al.}
\newblock \bibinfo{title}{Sunlight-powered sustained flight of an ultralight
  micro aerial vehicle}.
\newblock \emph{\bibinfo{journal}{Nature}} \textbf{\bibinfo{volume}{631}},
  \bibinfo{pages}{537--543} (\bibinfo{year}{2024}).

\bibitem{farrell2018review}
\bibinfo{author}{Farrell~Helbling, E.} \& \bibinfo{author}{Wood, R.~J.}
\newblock \bibinfo{title}{A review of propulsion, power, and control
  architectures for insect-scale flapping-wing vehicles}.
\newblock \emph{\bibinfo{journal}{Applied Mechanics Reviews}}
  \textbf{\bibinfo{volume}{70}}, \bibinfo{pages}{010801}
  (\bibinfo{year}{2018}).

\bibitem{sui2025untethered}
\bibinfo{author}{Sui, F.} \emph{et~al.}
\newblock \bibinfo{title}{Untethered subcentimeter flying robots}.
\newblock \emph{\bibinfo{journal}{Science Advances}}
  \textbf{\bibinfo{volume}{11}}, \bibinfo{pages}{eads6858}
  (\bibinfo{year}{2025}).

\bibitem{jancik2025nano}
\bibinfo{author}{Jancik-Prochazkova, A.} \& \bibinfo{author}{Ariga, K.}
\newblock \bibinfo{title}{Nano-/microrobots for environmental remediation in
  the eyes of nanoarchitectonics: Toward engineering on a single-atomic scale}.
\newblock \emph{\bibinfo{journal}{Research}} \textbf{\bibinfo{volume}{8}},
  \bibinfo{pages}{0624} (\bibinfo{year}{2025}).

\bibitem{jiang2022control}
\bibinfo{author}{Jiang, J.}, \bibinfo{author}{Yang, Z.},
  \bibinfo{author}{Ferreira, A.} \& \bibinfo{author}{Zhang, L.}
\newblock \bibinfo{title}{Control and autonomy of microrobots: Recent progress
  and perspective}.
\newblock \emph{\bibinfo{journal}{Advanced Intelligent Systems}}
  \textbf{\bibinfo{volume}{4}}, \bibinfo{pages}{2100279}
  (\bibinfo{year}{2022}).

\bibitem{Chen2025NoiseTolerantHMI}
\bibinfo{author}{Chen, X.}, \bibinfo{author}{Lou, Z.}, \bibinfo{author}{Gao,
  X.} \emph{et~al.}
\newblock \bibinfo{title}{A noise-tolerant human--machine interface based on
  deep learning-enhanced wearable sensors}.
\newblock \emph{\bibinfo{journal}{Nature Sensors}}  (\bibinfo{year}{2025}).
\newblock \urlprefix\url{https://doi.org/10.1038/s44460-025-00001-3}.
\newblock \bibinfo{note}{Published: 17 November 2025}.

\bibitem{ates2022end}
\bibinfo{author}{Ates, H.~C.} \emph{et~al.}
\newblock \bibinfo{title}{End-to-end design of wearable sensors}.
\newblock \emph{\bibinfo{journal}{Nature Reviews Materials}}
  \textbf{\bibinfo{volume}{7}}, \bibinfo{pages}{887--907}
  (\bibinfo{year}{2022}).

\bibitem{kowaltschek2012lessons}
\bibinfo{author}{Kowaltschek, S.}
\newblock \bibinfo{title}{Lessons learnt from the sireus mems detector
  evaluation}.
\newblock In \emph{\bibinfo{booktitle}{Proc. 6th ESA/ESTEC Workshop Avionics
  Data, Control Softw. Syst.}}, \bibinfo{pages}{1--16} (\bibinfo{year}{2012}).

\bibitem{haavisto1983thin}
\bibinfo{author}{Haavisto, J.~R.}
\newblock \bibinfo{title}{Thin-film waveguides for inertial sensors}.
\newblock In \emph{\bibinfo{booktitle}{Fiber Optic and Laser Sensors I}}, vol.
  \bibinfo{volume}{412}, \bibinfo{pages}{221--228}
  (\bibinfo{organization}{SPIE}, \bibinfo{year}{1983}).

\bibitem{post1967sagnac}
\bibinfo{author}{Post, E.~J.}
\newblock \bibinfo{title}{Sagnac effect}.
\newblock \emph{\bibinfo{journal}{Reviews of Modern Physics}}
  \textbf{\bibinfo{volume}{39}}, \bibinfo{pages}{475} (\bibinfo{year}{1967}).

\bibitem{song2023advanced}
\bibinfo{author}{Song, N.}, \bibinfo{author}{Xu, X.}, \bibinfo{author}{Zhang,
  Z.}, \bibinfo{author}{Gao, F.} \& \bibinfo{author}{Wang, X.}
\newblock \bibinfo{title}{Advanced interferometric fiber optic gyroscope for
  inertial sensing: A review}.
\newblock \emph{\bibinfo{journal}{Journal of lightwave technology}}
  \textbf{\bibinfo{volume}{41}}, \bibinfo{pages}{4023--4034}
  (\bibinfo{year}{2023}).

\bibitem{bogaerts2012silicon}
\bibinfo{author}{Bogaerts, W.} \emph{et~al.}
\newblock \bibinfo{title}{Silicon microring resonators}.
\newblock \emph{\bibinfo{journal}{Laser \& photonics reviews}}
  \textbf{\bibinfo{volume}{6}}, \bibinfo{pages}{47--73} (\bibinfo{year}{2012}).

\bibitem{jiang2020whispering}
\bibinfo{author}{Jiang, X.}, \bibinfo{author}{Qavi, A.~J.},
  \bibinfo{author}{Huang, S.~H.} \& \bibinfo{author}{Yang, L.}
\newblock \bibinfo{title}{Whispering-gallery sensors}.
\newblock \emph{\bibinfo{journal}{Matter}} \textbf{\bibinfo{volume}{3}},
  \bibinfo{pages}{371--392} (\bibinfo{year}{2020}).

\bibitem{lefevre1993fiber}
\bibinfo{author}{Lefevre, H.~C.}
\newblock \bibinfo{title}{Fiber optic gyroscopes}.
\newblock In \emph{\bibinfo{booktitle}{Optical Fiber Sensors}},
  \bibinfo{pages}{69--72} (\bibinfo{organization}{Springer},
  \bibinfo{year}{1993}).

\bibitem{liang2017resonant}
\bibinfo{author}{Liang, W.} \emph{et~al.}
\newblock \bibinfo{title}{Resonant microphotonic gyroscope}.
\newblock \emph{\bibinfo{journal}{Optica}} \textbf{\bibinfo{volume}{4}},
  \bibinfo{pages}{114--117} (\bibinfo{year}{2017}).

\bibitem{liu2025chip}
\bibinfo{author}{Liu, S.} \emph{et~al.}
\newblock \bibinfo{title}{Chip-scale integrated optical gyroscope based on a
  multi-mode co-detection technique}.
\newblock \emph{\bibinfo{journal}{Photonics Research}}
  \textbf{\bibinfo{volume}{13}}, \bibinfo{pages}{319--329}
  (\bibinfo{year}{2025}).

\bibitem{li2017microresonator}
\bibinfo{author}{Li, J.}, \bibinfo{author}{Suh, M.-G.} \&
  \bibinfo{author}{Vahala, K.}
\newblock \bibinfo{title}{Microresonator brillouin gyroscope}.
\newblock \emph{\bibinfo{journal}{Optica}} \textbf{\bibinfo{volume}{4}},
  \bibinfo{pages}{346--348} (\bibinfo{year}{2017}).

\bibitem{lai2020earth}
\bibinfo{author}{Lai, Y.-H.} \emph{et~al.}
\newblock \bibinfo{title}{Earth rotation measured by a chip-scale ring laser
  gyroscope}.
\newblock \emph{\bibinfo{journal}{Nature Photonics}}
  \textbf{\bibinfo{volume}{14}}, \bibinfo{pages}{345--349}
  (\bibinfo{year}{2020}).

\bibitem{ge2026integrated}
\bibinfo{author}{Ge, J.} \emph{et~al.}
\newblock \bibinfo{title}{Integrated microresonator {Brillouin} gyroscope
  demonstrating 37 deg/h rotation measurements}.
\newblock \emph{\bibinfo{journal}{Optics Letters}}
  \textbf{\bibinfo{volume}{51}}, \bibinfo{pages}{5854--5857}
  (\bibinfo{year}{2026}).

\bibitem{suzuki2000monolithically}
\bibinfo{author}{Suzuki, K.}, \bibinfo{author}{Takiguchi, K.} \&
  \bibinfo{author}{Hotate, K.}
\newblock \bibinfo{title}{Monolithically integrated resonator microoptic gyro
  on silica planar lightwave circuit}.
\newblock \emph{\bibinfo{journal}{Journal of Lightwave Technology}}
  \textbf{\bibinfo{volume}{18}}, \bibinfo{pages}{66--72}
  (\bibinfo{year}{2000}).

\bibitem{wang2015resonator}
\bibinfo{author}{Wang, J.}, \bibinfo{author}{Feng, L.}, \bibinfo{author}{Tang,
  Y.} \& \bibinfo{author}{Zhi, Y.}
\newblock \bibinfo{title}{Resonator integrated optic gyro employing trapezoidal
  phase modulation technique}.
\newblock \emph{\bibinfo{journal}{Optics letters}}
  \textbf{\bibinfo{volume}{40}}, \bibinfo{pages}{155--158}
  (\bibinfo{year}{2015}).

\bibitem{huang2021rotation}
\bibinfo{author}{Huang, J.-H.}, \bibinfo{author}{Duan, X.-Y.} \&
  \bibinfo{author}{Hu, X.-Y.}
\newblock \bibinfo{title}{Amplification of rotation velocity using weak
  measurements in {Sagnac}'s interferometer}.
\newblock \emph{\bibinfo{journal}{The European Physical Journal D}}
  \textbf{\bibinfo{volume}{75}}, \bibinfo{pages}{114} (\bibinfo{year}{2021}).

\bibitem{song2021enhanced}
\bibinfo{author}{Song, M.} \emph{et~al.}
\newblock \bibinfo{title}{Enhanced on-chip phase measurement by inverse weak
  value amplification}.
\newblock \emph{\bibinfo{journal}{Nature Communications}}
  \textbf{\bibinfo{volume}{12}}, \bibinfo{pages}{6247} (\bibinfo{year}{2021}).

\bibitem{yanik2025development}
\bibinfo{author}{Yanik, K.}, \bibinfo{author}{Song, M.}, \bibinfo{author}{Mei,
  Y.}, \bibinfo{author}{Cardenas, J.} \& \bibinfo{author}{Jordan, A.~N.}
\newblock \bibinfo{title}{Development of a chip-scale optical gyroscope with
  weak measurement amplification readout}.
\newblock \emph{\bibinfo{journal}{Opt. Express}} \textbf{\bibinfo{volume}{34}},
  \bibinfo{pages}{34898--34917} (\bibinfo{year}{2026}).

\bibitem{malykin2014sagnac}
\bibinfo{author}{Malykin, G.~B.}
\newblock \bibinfo{title}{Sagnac effect in ring lasers and ring resonators. how
  does the refractive index of the optical medium influence the sensitivity to
  rotation?}
\newblock \emph{\bibinfo{journal}{Physics-Uspekhi}}
  \textbf{\bibinfo{volume}{57}}, \bibinfo{pages}{714} (\bibinfo{year}{2014}).

\bibitem{volpe2016effective}
\bibinfo{author}{Volpe, G.} \& \bibinfo{author}{Wehr, J.}
\newblock \bibinfo{title}{Effective drifts in dynamical systems with
  multiplicative noise: a review of recent progress}.
\newblock \emph{\bibinfo{journal}{Reports on Progress in Physics}}
  \textbf{\bibinfo{volume}{79}}, \bibinfo{pages}{053901}
  (\bibinfo{year}{2016}).

\bibitem{aharonov1988result}
\bibinfo{author}{Aharonov, Y.}, \bibinfo{author}{Albert, D.~Z.} \&
  \bibinfo{author}{Vaidman, L.}
\newblock \bibinfo{title}{How the result of a measurement of a component of the
  spin of a spin-1/2 particle can turn out to be 100}.
\newblock \emph{\bibinfo{journal}{Physical review letters}}
  \textbf{\bibinfo{volume}{60}}, \bibinfo{pages}{1351} (\bibinfo{year}{1988}).

\bibitem{grant2022chip}
\bibinfo{author}{Grant, M.~J.} \emph{et~al.}
\newblock \bibinfo{title}{Chip-scale gyroscope using silicon-nitride waveguide
  resonator with a q factor of 100 million}.
\newblock In \emph{\bibinfo{booktitle}{Optical and Quantum Sensing and
  Precision Metrology II}}, vol. \bibinfo{volume}{12016},
  \bibinfo{pages}{175--185} (\bibinfo{organization}{SPIE},
  \bibinfo{year}{2022}).

\bibitem{zawada2025chip}
\bibinfo{author}{Zawada, A.~N.} \emph{et~al.}
\newblock \bibinfo{title}{Chip-scale resonant optical gyroscope with near
  earth-rate sensitivity}.
\newblock \emph{\bibinfo{journal}{IEEE Sensors Journal}}
  (\bibinfo{year}{2025}).

\bibitem{zawada2024passive}
\bibinfo{author}{Zawada, A.~N.} \emph{et~al.}
\newblock \bibinfo{title}{Passive chip-scale resonant optical gyroscope with
  sub-20-deg/hour/{$\sqrt{\mathrm{Hz}}$} performance}.
\newblock In \emph{\bibinfo{booktitle}{Quantum Sensing, Imaging, and Precision
  Metrology II}}, vol. \bibinfo{volume}{12912}, \bibinfo{pages}{101--107}
  (\bibinfo{organization}{SPIE}, \bibinfo{year}{2024}).

\bibitem{feng2021resonant}
\bibinfo{author}{Feng, C.} \emph{et~al.}
\newblock \bibinfo{title}{Resonant integrated optical gyroscope based on si3n4
  waveguide ring resonator}.
\newblock \emph{\bibinfo{journal}{Optics Express}}
  \textbf{\bibinfo{volume}{29}}, \bibinfo{pages}{43875--43884}
  (\bibinfo{year}{2021}).

\bibitem{zhang2017single}
\bibinfo{author}{Zhang, J.}, \bibinfo{author}{Ma, H.}, \bibinfo{author}{Li, H.}
  \& \bibinfo{author}{Jin, Z.}
\newblock \bibinfo{title}{Single-polarization fiber-pigtailed high-finesse
  silica waveguide ring resonator for a resonant micro-optic gyroscope}.
\newblock \emph{\bibinfo{journal}{Optics letters}}
  \textbf{\bibinfo{volume}{42}}, \bibinfo{pages}{3658--3661}
  (\bibinfo{year}{2017}).

\bibitem{hu2026ultra}
\bibinfo{author}{Hu, J.}, \bibinfo{author}{Liu, S.}, \bibinfo{author}{Li, B.},
  \bibinfo{author}{Yong, Y.} \& \bibinfo{author}{Ma, H.}
\newblock \bibinfo{title}{Ultra-high stability chip-scale optical gyroscope}.
\newblock \emph{\bibinfo{journal}{Journal of Lightwave Technology}}
  \textbf{\bibinfo{volume}{44}}, \bibinfo{pages}{2565--2570}
  (\bibinfo{year}{2026}).

\bibitem{wang2025broadband}
\bibinfo{author}{Wang, Y.} \emph{et~al.}
\newblock \bibinfo{title}{Broadband source-driven resonant integrated optical
  gyroscope using a low modulation frequency}.
\newblock \emph{\bibinfo{journal}{Optics Express}}
  \textbf{\bibinfo{volume}{33}}, \bibinfo{pages}{36172--36181}
  (\bibinfo{year}{2025}).

\bibitem{feng2022improving}
\bibinfo{author}{Feng, C.}, \bibinfo{author}{Zhang, Y.}, \bibinfo{author}{Ma,
  H.}, \bibinfo{author}{Li, H.} \& \bibinfo{author}{Feng, L.}
\newblock \bibinfo{title}{Improving long-term temperature bias stability of an
  integrated optical gyroscope employing a si3n4 resonator}.
\newblock \emph{\bibinfo{journal}{Photonics Research}}
  \textbf{\bibinfo{volume}{10}}, \bibinfo{pages}{1661--1668}
  (\bibinfo{year}{2022}).

\bibitem{lai2019observation}
\bibinfo{author}{Lai, Y.-H.}, \bibinfo{author}{Lu, Y.-K.},
  \bibinfo{author}{Suh, M.-G.}, \bibinfo{author}{Yuan, Z.} \&
  \bibinfo{author}{Vahala, K.}
\newblock \bibinfo{title}{Observation of the exceptional-point-enhanced sagnac
  effect}.
\newblock \emph{\bibinfo{journal}{Nature}} \textbf{\bibinfo{volume}{576}},
  \bibinfo{pages}{65--69} (\bibinfo{year}{2019}).

\bibitem{feng2024interferometric}
\bibinfo{author}{Feng, C.} \emph{et~al.}
\newblock \bibinfo{title}{Interferometric integrated optical gyroscope based on
  silicon nitride waveguide ring}.
\newblock \emph{\bibinfo{journal}{IEEE Sensors Journal}}
  \textbf{\bibinfo{volume}{24}}, \bibinfo{pages}{23781--23788}
  (\bibinfo{year}{2024}).

\bibitem{gundavarapu2018interferometric}
\bibinfo{author}{Gundavarapu, S.} \emph{et~al.}
\newblock \bibinfo{title}{Interferometric optical gyroscope based on an
  integrated {Si$_3$N$_4$} low-loss waveguide coil}.
\newblock \emph{\bibinfo{journal}{Journal of Lightwave Technology}}
  \textbf{\bibinfo{volume}{36}}, \bibinfo{pages}{1185--1191}
  (\bibinfo{year}{2018}).

\bibitem{wang2016suppression}
\bibinfo{author}{Wang, J.}, \bibinfo{author}{Feng, L.}, \bibinfo{author}{Wang,
  Q.}, \bibinfo{author}{Jiao, H.} \& \bibinfo{author}{Wang, X.}
\newblock \bibinfo{title}{Suppression of backreflection error in resonator
  integrated optic gyro by the phase difference traversal method}.
\newblock \emph{\bibinfo{journal}{Optics letters}}
  \textbf{\bibinfo{volume}{41}}, \bibinfo{pages}{1586--1589}
  (\bibinfo{year}{2016}).

\bibitem{liu2020interferometric}
\bibinfo{author}{Liu, D.} \emph{et~al.}
\newblock \bibinfo{title}{Interferometric optical gyroscope based on an
  integrated silica waveguide coil with low loss}.
\newblock \emph{\bibinfo{journal}{Optics Express}}
  \textbf{\bibinfo{volume}{28}}, \bibinfo{pages}{15718--15730}
  (\bibinfo{year}{2020}).

\bibitem{li2022frequency}
\bibinfo{author}{Li, H.} \emph{et~al.}
\newblock \bibinfo{title}{Frequency spectrum separation method of suppressing
  backward-light-related errors for resonant integrated optical gyroscope}.
\newblock \emph{\bibinfo{journal}{Journal of Lightwave Technology}}
  \textbf{\bibinfo{volume}{40}}, \bibinfo{pages}{1188--1194}
  (\bibinfo{year}{2022}).

\bibitem{nan2024triple}
\bibinfo{author}{Nan, C.} \emph{et~al.}
\newblock \bibinfo{title}{Triple closed-loop resonant micro-optical gyroscope
  based on optoelectronic hybrid feedback}.
\newblock \emph{\bibinfo{journal}{Optical Engineering}}
  \textbf{\bibinfo{volume}{63}}, \bibinfo{pages}{054116--054116}
  (\bibinfo{year}{2024}).

\end{thebibliography}
\bibliographystyle{naturemag}
\section*{Methods}
\subsection*{Theoretical framework for the two-chain decoupling architecture}
We propose a two-chain decoupling architecture with weak value amplification. To estimate a parameter associated with the coupling between the QS(quantum system) and MA(measurement apparatus), the QS operator $\hat{A}$ is coupled to the MA operator $\hat{P}$ through the coupling Hamiltonian $H=f(t) \hat{P} \hat{A}$, where $f(t)$ is a coupling function with finite support that satisfies $\int f(t) d t=$ $g \ll 1$. 

Here, the QS is a two-level which-path MZI system with eigenstates $|u\rangle$ and $|d\rangle$ to represent the up-arm and down-arm paths, and the corresponding eigenvalues are +1 and -1. MA operator is $\hat{n}$.

The initial states of QS and MA are,
respectively,
$\left|\psi_i\right\rangle=\frac{1}{\sqrt{2}}
(e^{i\phi/2}|u\rangle+ie^{-i\phi/2}|d\rangle)$
and $\left|\alpha\right\rangle$.
After the interaction mediated by the dual-ring resonators, the QS and MA evolve
into the joint state
\begin{equation}
\left|\Psi\right\rangle
= T e^{-i\theta\hat{A}\hat{n}}
\left|\psi_i\right\rangle\left|\alpha\right\rangle
= \frac{1}{\sqrt{2}}
\left(
e^{i\phi\hat{n}/2}|u\rangle\left|\varphi_{u}\right\rangle
+ i e^{-i\phi\hat{n}/2}|d\rangle\left|\varphi_{d}\right\rangle
\right),
\label{eq:after coupling}
\end{equation}
where
$\hat{A} = \left|u \right \rangle \left \langle u \right| -\left|d \right \rangle \left \langle d \right|$, $\left|\varphi_{u,d}\right\rangle
= T e^{\mp i \theta \hat{n}}\left|\alpha\right\rangle$.
Here, $T$ and $\theta$ characterize the amplitude and
phase shift imparted by the resonator--waveguide coupling, respectively.

Under the initially on-resonance condition and a weak Sagnac phase perturbation,
$\Delta\phi_{\mathrm{Sagnac}} \ll 1$, the transmission amplitude and phase
responses of the micro-resonator can be approximated as
\begin{equation}
\begin{aligned}
T^2 &=
\frac{(r-a)^2 + r a \left(\Delta \phi_{\mathrm{Sagnac}}/2\right)^2}
{(1-r a)^2 + r a \left(\Delta \phi_{\mathrm{Sagnac}}/2\right)^2}, \\
\theta &\simeq
\frac{1}{2}\, G_r \, \Delta\phi_{\mathrm{Sagnac}} ,
\end{aligned}
\end{equation}
where the constant $\pi$ term in $\theta$ has been omitted, as it corresponds to
a global phase factor and does not affect the weak-measurement dynamics. Here,
\begin{equation}
G_r \equiv 1 + \frac{r}{a-r} + \frac{r a}{1-r a}
\end{equation}
denotes the resonant phase-enhancement factor of the micro-resonator, and $r$ and
$a$ represent the field self-coupling coefficient and the round-trip attenuation,
respectively. {This approximation requires near-resonant operation} and {coupling and loss perturbations satisfying $|\delta a-\delta r|\ll|a-r|$, a condition that can generally be satisfied for small perturbations;} the {full} validity {conditions and perturbation analysis} are provided in {the} Supplementary Information.

Notably, the amplitude response $T$ is insensitive to the Sagnac phase perturbation to first order, whereas the phase response $\theta$ exhibits a linear dependence on $\Delta\phi_{\mathrm{Sagnac}}$ with an enhancement factor
$G_r$. In the low-loss limit $a \to 1$, the transmission amplitude approaches unity, indicating that the resonator primarily acts as a phase-selective enhancement element for rotation-induced nonreciprocal perturbations rather than an amplitude modulator.

When a post-selection into $\left|\psi_f\right\rangle=\frac{1}{\sqrt{2}}(|u\rangle - i|d\rangle)$ is made on the QS, the MA state is collapsed into
\begin{equation}
\left|\varphi\right\rangle =\left\langle \psi_f\left|Te^{-i \theta \hat{A} \hat{n}}\right| \psi_i\right\rangle|\alpha\rangle \\
 =\frac{1}{2}e^{i\phi\hat{n}/2}\left|\varphi_{u}\right\rangle-\frac{1}{2} e^{-i\phi\hat{n}/2}\left|\varphi_{d}\right\rangle.
    \label{eq:post-selection}
\end{equation}

Under the condition
\begin{equation}
   |\theta| \ll \left|\frac{1}{A_w}\right|,
\end{equation}
where
\begin{equation}
A_w=\frac{\left\langle\psi_f|\mathbf{A}| \psi_i\right\rangle}
{\left\langle\psi_f \mid \psi_i\right\rangle}
= i\cot(\phi/2) \approx \frac{2i}{\phi}
\end{equation}
is the so-called weak value, Eq.~\ref{eq:post-selection} can be approximated as
\begin{equation}
|\varphi\rangle
= \langle \psi_f \mid \psi_i\rangle\,
T e^{-i \theta A_w \hat{n}}|\alpha\rangle .
\end{equation}
This expression explicitly shows that the small phase shift $\theta$ is
effectively amplified by the weak value $A_w$ in the post-selected meter state.
Accordingly, the mean photon number $\langle n\rangle$ can be calculated as
\begin{align}
\langle n\rangle
&= T^2 |\alpha|^2 \sin^2\!\left(\phi/2\right)
\left[1 + 2|A_w|\theta\right] \nonumber\\
&= K\left[1 + |A_w|
G_r\Delta\phi_{\mathrm{Sagnac}}\right],
\label{eq:n_A}
\end{align}
where $K \equiv T^2 |\alpha|^2 \sin^2\!\left(\phi/2\right)$ is independent of the Sagnac phase perturbation and can be normalized by the zero-rotation reference. Notably, $|A_w|$ amplifies only the resonator-enhanced Sagnac-phase term and does not act on the overall prefactor $K$, and therefore does not amplify other perturbations that are non-isomorphic to the Sagnac phase.

\begingroup\color{black}
\subsection*{Sensing-chip design and fabrication}
The gyroscope sensing chip was implemented in a Si$_3$N$_4$ multi-project-wafer process. The Si$_3$N$_4$ core is $400~\mathrm{nm}$ thick, with top and bottom oxide claddings of $4.1~\mu\mathrm{m}$ and $3.37~\mu\mathrm{m}$, respectively. The dual-ring sensing core comprises two nominally identical racetrack micro-ring resonators embedded in the two arms of the MZI. The resonators use $3~\mu\mathrm{m}$-wide multimode waveguides. Each ring has a perimeter of approximately $4.4~\mathrm{mm}$ and encloses an area of $0.8~\mathrm{mm}^2$. Euler bends with a maximum radius of $R_{\max}=1000~\mu\mathrm{m}$ and a minimum radius of $R_{\min}=150~\mu\mathrm{m}$ provide adiabatic fundamental-mode routing within the compact layout. Thermo-optic heaters independently tune the resonance alignment of the two rings and the static phase bias of the MZI. 
 Through-port spectra were measured to extract the loaded Q factors ($\sim$ $2.4\times10^6$) and the FSRs of the two rings. Resonance assignment, fitting details and the complete statistics are provided in Supplementary Information.

\endgroup

\begingroup\color{black}
\subsection*{Chip-scale SIL source and measurement configuration}
The chip-scale SIL source was assembled by butt-coupling a commercial DFB laser to a low-loss Si$_3$N$_4$ feedback chip. The DFB die employed a high-reflectivity facet and an anti-reflection-coated coupling facet. The feedback chip is on the same platform as sensing chip. A multimode interferometer then directed the field to a multimode-waveguide add--drop ring and a Sagnac-loop reflector, which returned resonant optical feedback to the DFB laser. The simulated coupling coefficient of the MRR was approximately 0.02. The ring had an $87$-mm perimeter and a measured FSR of approximately $0.014~\mathrm{nm}$ ($1.75~\mathrm{GHz}$). A resonance near 1547.8~nm had a fitted linewidth of $1.1~\mathrm{pm}$, corresponding to a loaded quality factor of approximately $1.4\times10^6$.

The SIL source and the gyroscope sensing chip were mounted on separate thermoelectric coolers. The SIL output was connected to the gyroscope through the optical path shown in Fig.~\ref{fig:SIL}a. The ring heaters and MZI bias were set before acquisition; no dynamic phase modulation or resonance-frequency-tracking servo was applied during the rotation measurement. The stage was driven at $0.1~\mathrm{Hz}$ for four angular-rate amplitudes. A separate two-hour zero-input record was acquired for static-noise characterization.

Here, ``open-loop operation without modulation'' refers to direct intensity-based angular-rate readout without angular-rate feedback, resonance-frequency tracking, or time-dependent modulation/demodulation. The thermo-optic phase shifters are used only for static resonance alignment and bias setting before measurement, and a TEC stabilizes the chip temperature during measurement.

\endgroup

\subsection*{Gyroscope Performance Characterization}

\begingroup\color{black}
Static performance was evaluated from continuous two-hour zero-input records using the matching angular-rate conversion described in Results. All static records were sampled at $10~\mathrm{kHz}$, with $7200~\mathrm{s}$ and $72{,}000{,}000$ samples analysed for each record. Only the constant mean was removed; no filtering, smoothing, pre-averaging or linear detrending was applied. For an averaging time $\tau$, non-overlapping interval means $\overline{\Omega}_i$ were used to calculate
\begin{equation}
\sigma_\Omega(\tau)=\sqrt{\frac{1}{2(M-1)}\sum_{i=1}^{M-1}\left(\overline{\Omega}_{i+1}-\overline{\Omega}_i\right)^2},
\end{equation}
where $M$ is the number of complete intervals. The non-overlapping Allan deviation was calculated using the full-resolution samples, with any incomplete trailing interval excluded. ARW was obtained from the fitted $\tau^{-1/2}$ region, with $\mathrm{ARW}=\sigma_\Omega(\tau)\sqrt{\tau}$ when $\tau$ is expressed in hours and $\sigma_\Omega$ in $^\circ/\mathrm{h}$. BI is reported as the minimum Allan deviation. Approximate pointwise conditional $95\%$ confidence intervals were estimated from the $2.5$th and $97.5$th percentiles of $10{,}000$ circular time-block bootstrap replicates, using a primary block length of $360~\mathrm{s}$.

\endgroup

\clearpage
\section*{Extended data}
\setcounter{figure}{0}

\begin{figure*}[htbp]
  \centering
  \includegraphics[width=1\linewidth]{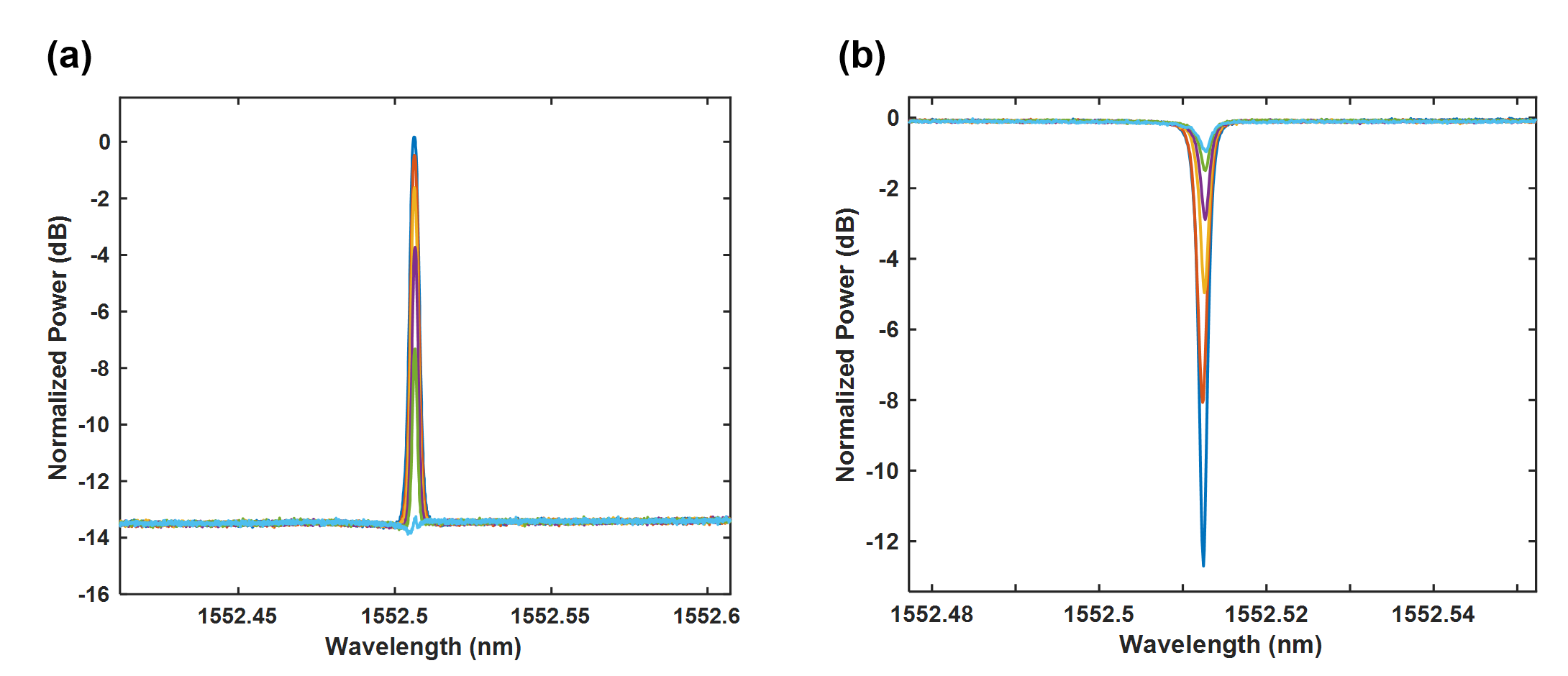}
  \begingroup
  \makeatletter
  \renewcommand{\fnum@figure}{}
  \makeatother
    \caption{\textbf{Extended Data Fig. 1. Experimental intensity-based spectral response to different non-reciprocal phase bias.}
A sequence of small phase biases is applied by the phase modulator on one ring, producing a non-reciprocal phase bias between the two arms of this architecture, and the output spectra are measured at the (a) bar and (b) cross ports. Curves in different colors correspond to different applied phase biases. Unlike conventional resonator readout, where a phase perturbation mainly manifests as a resonance frequency shift, the present architecture maps non-reciprocal phase predominantly into intensity change of the resonance peak.}
  \label{fig:ED1}
  \endgroup
\end{figure*}

\begin{figure*}[htbp]
  \centering
  \includegraphics[width=1\linewidth]{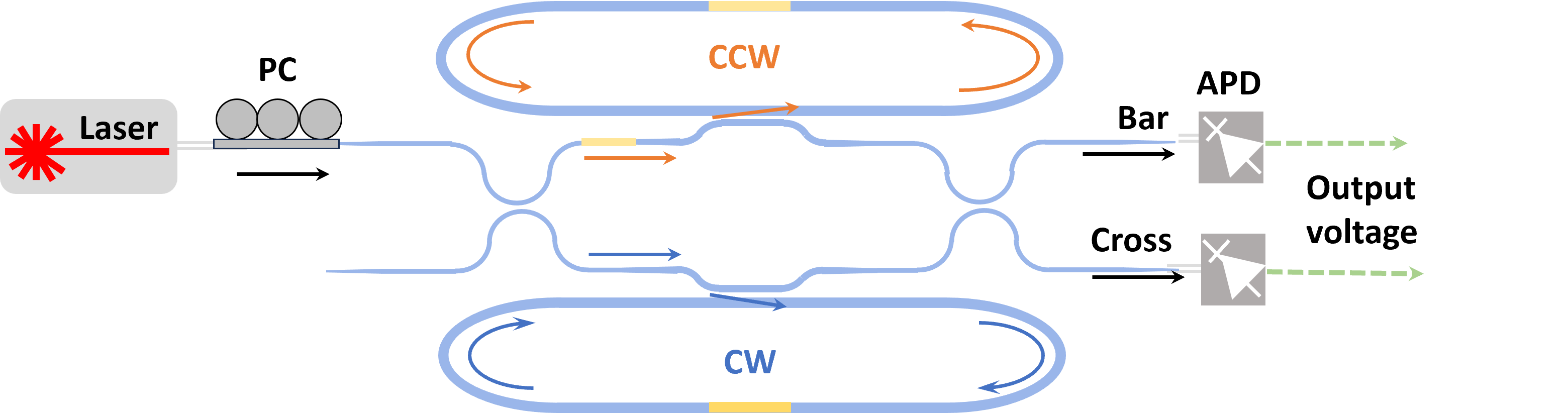}
  \begingroup
  \makeatletter
  \renewcommand{\fnum@figure}{}
  \makeatother
\caption{\textbf{Extended Data Fig. 2. Experimental measurement setup.}
A narrow-linewidth single-frequency laser is used as the optical source and injected into the system. A polarization controller (PC) is employed to ensure predominantly transverse-electric (TE) fundamental-mode excitation in the chip. The bar and cross ports are connected to avalanche photodiode (APD) detectors, which convert the optical responses into electrical voltage signals for readout.}
 \label{fig:ED2}
  \endgroup
\end{figure*}

\begin{figure}[htbp]
\centering
\captionsetup{font=small,labelfont=bf}
\includegraphics[width=0.90\linewidth]{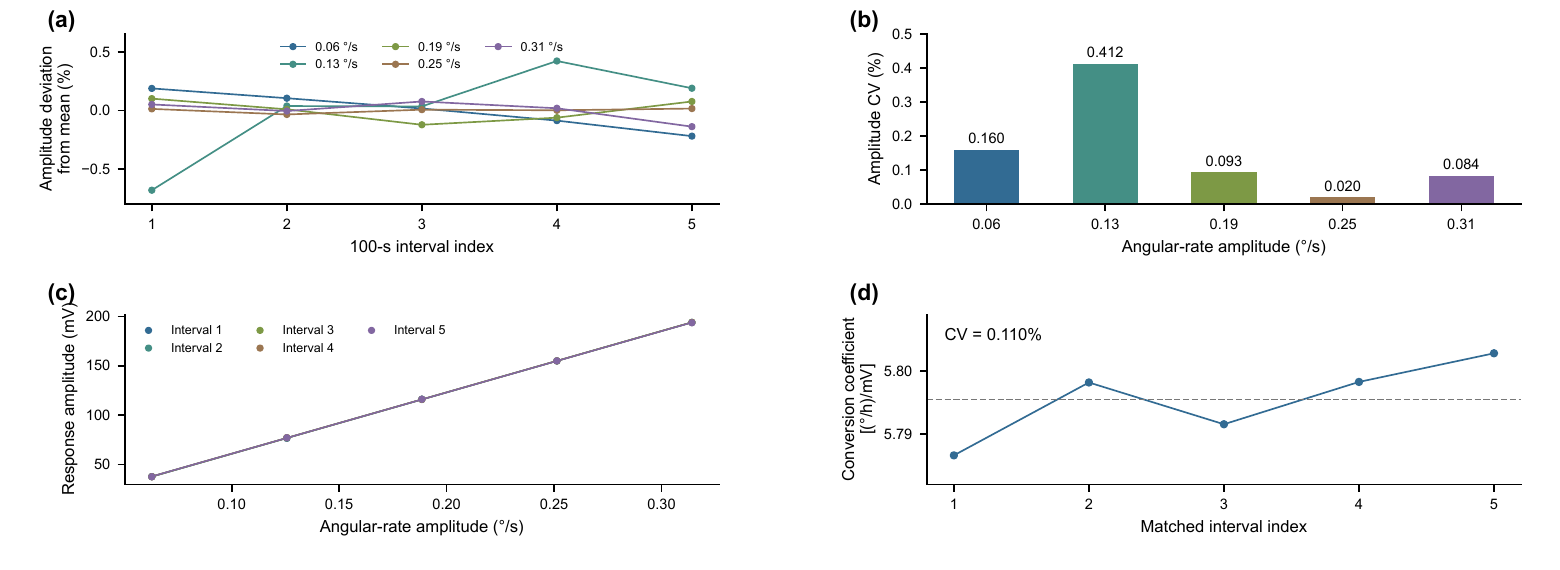}
\captionsetup{font=small}
\renewcommand{\figurename}{Extended Data Fig.}
\renewcommand{\thefigure}{3}
\caption{{\textbf{Temporal repeatability of the dynamic response and consistency of cross-amplitude calibration.}
The five input angular-rate peak amplitudes are approximately $0.06$, $0.13$, $0.19$, $0.25$ and $0.31^\circ/\mathrm{s}$, all at $0.1~\mathrm{Hz}$. Each $500~\mathrm{s}$ record is divided into five consecutive $100~\mathrm{s}$ intervals. A sinusoidal fit at the known driving frequency, including a constant offset, is applied to each interval.
(a) Percentage deviations of the interval response amplitudes from their mean for each input condition.
(b) CVs of the interval response amplitudes for the five inputs are $0.160\%$, $0.412\%$, $0.093\%$, $0.020\%$ and $0.084\%$, respectively. CV is defined as the sample standard deviation divided by the absolute mean.
(c) Five cross-amplitude calibrations, each constructed from intervals with matching indices across the records. Points are fitted interval amplitudes; solid lines use the corresponding free-intercept linear-fit parameters. The overall calibration of the five mean responses gives $621.17~\mathrm{mV}/(^\circ/\mathrm{s})$, $R^2=0.999995$ and a maximum nonlinearity of $0.081\%$ (maximum absolute residual divided by the maximum fitted response).
(d) Voltage-to-angular-rate conversion coefficients from the five calibrations, with a mean of $5.79551~(^\circ/\mathrm{h})/\mathrm{mV}$, a sample standard deviation of $0.00639~(^\circ/\mathrm{h})/\mathrm{mV}$ and a CV of $0.110\%$. The dashed line marks the mean.}}
\label{fig:ed-calibration}
\end{figure}

\begin{figure}[htbp]
\centering
\captionsetup{font=small,labelfont=bf}
\includegraphics[width=\linewidth]{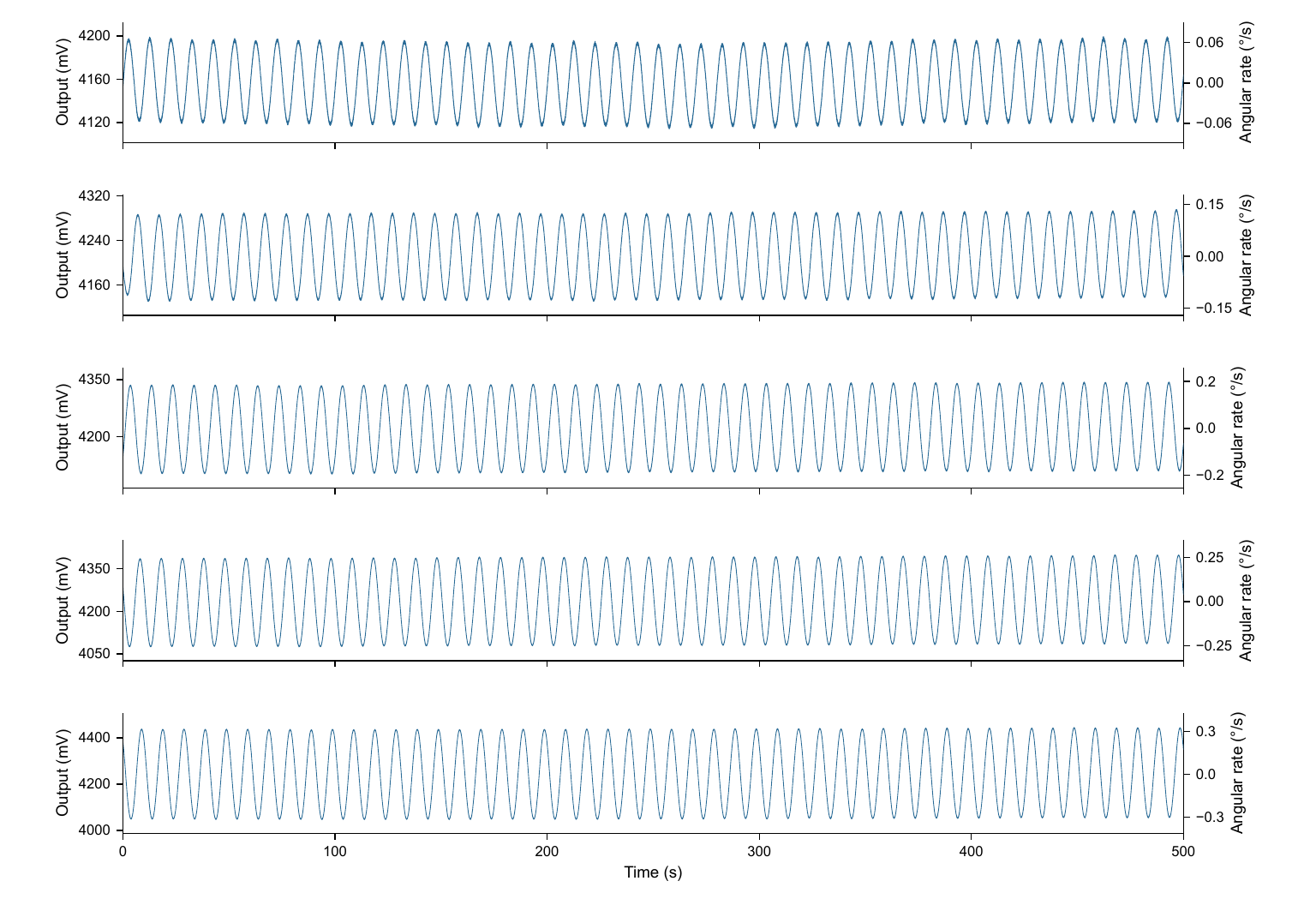}
\renewcommand{\figurename}{Extended Data Fig.}
\renewcommand{\thefigure}{4}
\caption{{\textbf{Five complete 500 s time-domain records used for dynamic calibration.}
The driving frequency is $0.1~\mathrm{Hz}$ in all cases, with angular-displacement amplitudes of $0.1^\circ$, $0.2^\circ$, $0.3^\circ$, $0.4^\circ$ and $0.5^\circ$ from top to bottom. Blue traces show the measured output over each complete $500~\mathrm{s}$ record. Each record is divided into five consecutive $100~\mathrm{s}$ intervals for sinusoidal-response amplitude extraction and interval calibration. Figure~\ref{fig:main}a shows the first $100~\mathrm{s}$ of each record.}}
\label{fig:supp-dynamic-500s}
\end{figure}

\begin{table*}[htbp]
\captionsetup{labelformat=empty,font=small,labelfont=bf}
\centering
\scriptsize
\setlength{\tabcolsep}{4pt}
\renewcommand{\arraystretch}{1.15}
\caption{Extended Data Table 1. Performance comparison of representative integrated optical gyroscopes.}
\label{tab:gyro_comparison}
\resizebox{\linewidth}{!}{
\begin{tabular}{llcccll}
\toprule
Ref. & Type & {\color{black}$Area$ ($\mathrm{cm}^2$)} & {\color{black}$BI$ ($^{\circ}/\mathrm{h}$)} & {\color{black}$ARW$ ($^{\circ}/\sqrt{\mathrm{h}}$)} & Resonator Type & Platform \\
\midrule\midrule
\textbf{This Work} & \textbf{PROG} & \textbf{0.008} & {\color{black}\textbf{0.79}} & {\color{black}\textbf{0.0014}} & \textbf{Waveguide resonator} & {\color{black}\textbf{Si$_3$N$_4$}} \\
Hajimiri 2018 \cite{khial2018nanophotonic} & PROG & 0.008 & 21600 & 650 & Waveguide resonator & Silicon \\
\shortstack[l]{Digonnet 2022\\\cite{grant2022chip}} & PROG & 0.35 & {\color{black}\shortstack[l]{Not reported (originally reported);\\ $\approx 600$ (minimum Allan deviation)}} & 1.3 & Waveguide resonator & {\color{black}Si$_3$N$_4$} \\
Maleki 2017 \cite{liang2017resonant} & PROG & {\color{black}0.39} & 3 & 0.02 & WGM microcavity & {\color{black}CaF$_2$} \\
{\color{black}Ma 2025a \cite{liu2025chip}} & PROG & {\color{black}0.66} & 1 & 0.8 & WGM microcavity & {\color{black}MgF$_2$} \\
{\color{black}\shortstack[l]{Digonnet 2025\\\cite{zawada2025chip}}} & {\color{black}PROG} & {\color{black}1.17} & {\color{black}\shortstack[l]{250 (originally reported);\\ $\approx 20$ (minimum Allan deviation)}} & {\color{black}0.11} & {\color{black}Waveguide resonator} & {\color{black}Si$_3$N$_4$} \\
{\color{black}\shortstack[l]{Digonnet 2024\\\cite{zawada2024passive}}} & {\color{black}PROG} & {\color{black}1.17} & {\color{black}\shortstack[l]{500 (originally reported);\\ $\approx 80$ (minimum Allan deviation)}} & {\color{black}0.28} & {\color{black}Waveguide resonator} & {\color{black}Si$_3$N$_4$} \\
Feng 2021 \cite{feng2021resonant} & PROG & 2.01 & 2448 & --- & Waveguide resonator & {\color{black}Si$_3$N$_4$} \\
{Ge 2026 \cite{ge2026integrated}} & {BROG} & {4.36} & {302} & {0.88} & {Waveguide resonator} & {Si$_3$N$_4$} \\
Ma 2017 \cite{zhang2017single} & PROG & 4.91 & 14.4 & --- & Waveguide resonator & {\color{black}SiO$_2$} \\
{\color{black}Ma 2026 \cite{hu2026ultra}} & PROG & 6.38 & 0.5 & 0.21 & WGM microcavity & {\color{black}---} \\
Zhang 2025 \cite{wang2025broadband} & PROG & 7.07 & 0.9 & 0.17 & WGM microcavity & {\color{black}MgF$_2$} \\
Feng 2022b \cite{feng2022improving} & PROG & 9.62 & 13.2 & --- & Waveguide resonator & {\color{black}Si$_3$N$_4$} \\
Vahala 2020 \cite{lai2020earth} & BROG & 10.18 & 3.6 & 0.068 & Wedge resonator & {\color{black}SiO$_2$} \\
\shortstack[l]{Vahala 2019\\\cite{lai2019observation}} & BROG & 10.18 & {\color{black}\shortstack[l]{Not reported (originally reported);\\ $\approx 7$ (minimum Allan deviation)}} & --- & Wedge resonator & {\color{black}SiO$_2$} \\
{\color{black}Li 2024 \cite{feng2024interferometric}} & IOG & 11.08 & 11.8 & --- & Coiled waveguide & {\color{black}Si$_3$N$_4$} \\
Blumenthal 2018 \cite{gundavarapu2018interferometric} & IOG & 12.56 & 58.7 & 8.52 & Coiled waveguide & {\color{black}Si$_3$N$_4$} \\
Zhi 2015 \cite{wang2015resonator} & PROG & 28.27 & 324 & --- & Waveguide resonator & {\color{black}SiO$_2$} \\
Feng 2016 \cite{wang2016suppression} & PROG & 28.27 & {\color{black}21.6--25.2} & --- & Waveguide resonator & {\color{black}SiO$_2$} \\
Feng 2020 \cite{liu2020interferometric} & IOG & 31.17 & 7.32 & 1.26 & Coiled waveguide & {\color{black}SiO$_2$} \\
Feng 2022a \cite{li2022frequency} & PROG & 78.54 & 0.3 & --- & Waveguide resonator & {\color{black}SiO$_2$} \\
{\color{black}Liu 2024 \cite{nan2024triple}} & PROG & {\color{black}113.1} & 0.926 & --- & Waveguide resonator & {\color{black}SiO$_2$} \\
\bottomrule
\end{tabular}}
\par\medskip
\begin{minipage}{\textwidth}
\normalsize
{For two-line BI entries, ``originally reported'' and ``minimum Allan deviation'' distinguish the source-reported quantity from the curve minimum. Values preceded by $\approx$ are approximate. In Refs.~\cite{zawada2025chip,zawada2024passive}, the reported quantity is drift, defined as the maximum of the Allan-deviation curve. For Ref.~\cite{zawada2024passive}, 500$^\circ$}\,{h$^{-1}$ is the drift after splice modifications, whereas approximately 80$^\circ$}\,{h$^{-1}$ is the minimum of Fig.~5 before those modifications.}
\par\smallskip
\end{minipage}
\end{table*}

\clearpage
\newpage
\section*{End Notes}
\subsection*{Acknowledgements}
This work was supported by the National Natural Science Foundation of China (No. 62471289, 62405184), Innovation Program for Quantum Science and Technology (No.2021ZD0300703), Shanghai Municipal Science and Technology Major Project (Grant No. 2019SHZDZX01), and the Natural Science Foundation of Shanghai (24ZR1431400).
\subsection*{Author Contributions}
Y.T. and J.H. conceived the idea, J.H. and G.Z. designed and supervised the project. Y.T.constructed the theoretical model, X.L., Y.T., Y. G. and L.Z. carried out the chip design and layout implementation for fabrication. Y.T. carried out the experiments with assistance from X.L., Y.G. and H.L., Y.T., J.H. and  Z.L. analyzed the data. Y.T. wrote the manuscript with contributions from all authors.
\subsection*{Declaration of Interests}
The authors declare no competing interests.

\end{document}